%% file: main.tex
\documentclass[%
reprint,
 amsmath,amssymb,
 aps,
prl]{revtex4-2}

\usepackage{braket}
\usepackage{hyperref}
\usepackage{graphicx}
\usepackage{dcolumn}
\usepackage{bm}
\usepackage[mathlines]{lineno}

\input{definitions}

\begin{document}

\preprint{APS/123-QED}

\title{First observation of quantum correlations in  $\ee\to X\DD$ \\and \C-even constrained \DD pairs}

\input{authorlist_2025-04-06}

\date{\today}

\begin{abstract}

The study of meson pairs produced with quantum correlations gives direct access to parameters that are challenging to measure in other systems. In this Letter, the existence of quantum correlations due to charge-conjugation symmetry \C are demonstrated in \DD pairs produced through the processes $\ee\to\DD$, $\ee \to \DSTD$, and $\ee \to \DSTDST$, where the lack of charge superscripts refers to an admixture of neutral-charm-meson particle and antiparticle states, using \mbox{7.13 \invfb} of $\ee$ collision data collected by the BESIII experiment between center-of-mass energies of \mbox{$4.13-4.23$ \GeV}. Processes with either \C-even or \C-odd constraints are identified and separated. A procedure is presented that harnesses the entangled production process to enable measurements of $\Dz$-meson hadronic parameters. This study provides the first confirmation of quantum correlations in  $\ee\to X \DD$ processes   and the first observation of a \C-even constrained \DD system.  The procedure is applied to measure $\deltaKpi$, the strong phase between the $\Dz\to \Km\pip$ and $\Dzb\to \Km\pip$ decay amplitudes, which results in the determination of $\deltaKpi=\left(192.8^{+11.0 + 1.9}_{-12.4 -2.4}\right)^\circ$.  The potential for measurements of other hadronic decay parameters and charm mixing with these and future datasets is also discussed.   
\end{abstract}

\maketitle


Quantum correlations induced by charge-conjugation
symmetry \C in the production of pairs of neutral $D$ mesons from
electron-positron collisions have been well-demonstrated
with data collected near energy threshold by the
CLEO and BESIII collaborations~\cite{CLEOQC,psippKspipiPRL,psippKspipiPRD,BESIIIK3Pi}. The production through a virtual photon and the intermediate vector $\psipp$ resonance both enforce the $\DDb$ final state to be \C-odd, where the lack of charge superscripts on the $\D$ mesons refers to an admixture of the neutral charm meson particle and antiparticle states. This constraint induces interference between the mass and charge-parity (\CP ) eigenstates of the $c \overline u$ system. The interference has been harnessed to determine hadronic parameters such as the
strong-phase difference between decays of $\Dz$ and $\Dzb$ mesons to the same final state. These hadronic parameters are critical inputs to the interpretation of mixing and \CP-violation measurements in the decays of charm and beauty quarks~\cite{LHCbBinFlip,LHCbKShh}.

The presence of such correlations for $\DDb$ pairs produced in higher energy $e^+e^-$ collisions, where additional particles may be produced, has been previously proposed~\cite{Goldhaber:1976fp,Xing:1996pn,Bondar:2010qs,Rama:2015pmr,Naik:2021rnv} but never examined experimentally. In these collisions, the \DD system can be  produced  as flavor-definite, or with either \C-odd or \C-even constraints, depending on the additional particles produced during the collision. When the accompanying particle system $X$ is composed of only a combination of photons and $\pi^0$ mesons, the expected \C eigenvalue of the $\DD$ pair is simply $\left(-1\right)^{n+1}$, where $n$ is the number of additional photons produced in the collision. Final states that are suppressed in a \C-odd initial state are enhanced in the \C-even case and vice-versa. Additionally, the interference  of $\DDb$ final states depends on the $\Dz$--\Dzb mixing parameters~\cite{LHCbBinFlip} $x$ and $y$ at first order for a \C-even constrained system, where $x$ and $y$ are enhanced relative to a flavor-definite state~\cite{Xing:1996pn,Bondar:2010qs}. In comparison, the dependence is quadratic in the \C-odd case. This dependence introduces the opportunity for measurements of charm mixing from $\ee$ collision data using a time-integrated method.

This Letter presents the study of quantum correlations in $\DD$ pairs produced in the processes $\ee\to\DD$, $\ee \to \DSTD$ \footnote{Charge conjugation is implied here and throughout the Letter.}, and $\ee \to \DSTDST$, and the first observation of \C-even constrained $\DD$ pairs with 7.13 \invfb of $\ee$ collision data collected by the BESIII experiment between center-of-mass energies of $4.13-4.23$ \GeV. The correlations are exploited to measure the strong phase $\deltaKpi$ between $\Dz\to \Km\pip$ and $\Dzb\to \Km\pip$ decays, which is described in detail in Ref.~\cite{CompanionPaper}. The potential for future measurements is also discussed. 
 
 The BESIII detector~\cite{BESIIIDetector} records the data from symmetric $\ee$ collisions provided by the BEPCII storage ring~\cite{BEPCII}. The cylindrical core of the BESIII detector covers $93\%$ of the full solid angle and consists of a helium-based multilayer drift chamber (MDC), a plastic scintillator time-of-flight system (TOF), and a CsI(Tl) electromagnetic calorimeter (EMC). All these detectors are enclosed in a superconducting solenoidal magnet providing a 1.0 T magnetic field. The solenoid is supported by an octagonal flux-return yoke with resistive plate counter muon-identification modules interleaved with steel. The end-cap TOF system was upgraded in 2015 using multigap resistive plate chamber technology. This upgrade occurred after the 1.1 $\text{fb}^{-1}$ of data was collected at the center-of-mass energy of 4.16 \GeV, but benefits all other data samples in this analysis.

 Simulated samples produced with a \textsc{geant4}-based~\cite{GEANT4}  (MC) package, which includes the geometric description of the BESIII detector and the detector response, are used to determine detection efficiencies and estimate backgrounds. The simulation models the beam-energy spread and initial-state radiation in the $\ee$ annihilations with the generator \textsc{kkmc}~\cite{KKMC}. The production of open-charm pairs, radiative production of charmonia, and continuum processes are also generated with \textsc{kkmc}. 
All particle decays are modeled with \textsc{evtgen}~\cite{EVTGEN,BESEVTGEN}, using branching fractions either taken from the Particle Data Group (PDG)~\cite{ParticleDataGroup:2024cfk} when available, or otherwise estimated with \textsc{lundcharm}~\cite{Chen:2000tv,Yang:2014vra}.  Final-state radiation from charged final-state particles is incorporated using the \textsc{photos} package~\cite{Barberio:1990ms}.  The simulation is produced without quantum correlations, allowing for a null-test comparison to the data.


The presence of quantum correlations is studied in five \mbox{$\ee\to X\DD$} production mechanisms expected to be \C-definite: \mbox{$\DD\;[\C=-1]$}, \mbox{$\DSTDG\;[\C=+1]$}, \mbox{$\DSTDP\;[\C=-1]$}, \mbox{$\DSTDSTEven\;[\C=+1]$}, and \mbox{$\DSTDSTOdd\;[\C=-1]$}. Quantum correlations are demonstrated in the production mechanisms of interest through the examination of the rates of $\DD$ pairs decaying to final states expected to be forbidden or enhanced under \C constraints with negligible dependence on poorly known external parameters. The examined final states are either comprised of two individual \CP-eigenstate \D decays~\footnote{The effects of direct $\CP$-violation in $\D$ decays can be safely neglected here, known to be of or below $\mathcal O(10^{-3})$ for the considered final states.}, or subject to exchange symmetries of the two \D mesons imposed on the overall \DD wavefunction. These rates are measured relative to the  $\DD\to K^+\pi^-\text{ vs. }K^-\pi^+$ final state, which has a rate that is affected negligibly by \C correlations. The measured rates of both forbidden and enhanced processes are interpreted together to confirm the presence of correlations in each production mechanism with expected eigenvalue $\C_\text{exp}$. The presence of the correlations is quantified through a coherence factor $\kappa$, defined by 
\begin{equation}\label{eq:kappadef}
    \hat {C}\ket{\DD}=\kappa C_\text{exp}\ket{\DD}-(1-\kappa) C_\text{exp}\ket{\DD}.
\end{equation}
 A sample which is completely coherent will have $\kappa=1$, and $0.5$ if exactly incoherent.

The \DD pairs decaying to the final states listed in Table~\ref{table:CPDTags} are reconstructed and assigned production-mechanism hypotheses using kinematic variables. Three variables using kinematic information of the reconstructed $\DD$ pair are used to isolate $\DD$ and $\DSTD$ events: 
\begin{equation}\label{eq:kinvars}
 \begin{aligned}
	\Emiss \equiv\Ecm&-E_{\DD}, \quad c^4\MMSq \equiv \Emiss^2-c^2\mathbf{p}_{\DD}^2,\\
		\text{and } \DStarMRec \equiv&\min\left(\Big|\MRec{\D}-m_{D^{*0}}\Big|,\Big|\MRec{\Db}-m_{D^{*0}}\Big|\right)\\
 \end{aligned}
 \end{equation}
where $\Ecm$ is the $\ee$ collision energy in the center-of-mass frame, $\mathbf{p}_{\DD}$ ($E_{\DD}$) is the three-momentum (energy) of the $\DD$ pair calculated under kinematic constraints to optimize resolution,  $m_{D^{*0}}$ corresponds to the known $D^{*0}$ mass~\cite{ParticleDataGroup:2024cfk}, and the recoil mass $\MRec{D}$ is defined as

\begin{equation}\label{eq:mRec}
c^2\MRec{D}\equiv\left[\left(\Ecm^2-\sqrt{c^2\mathbf{p}_{D}^2+c^4m_{D^{0}}^2}\right)^2-c^2\mathbf{p}_{D}^2\right]^{\frac 12}.
\end{equation}
$\Emiss$ is used to isolate $\ee\to\DD$ events, $\DStarMRec$ separates $\ee\to \DSTD$ from $\ee\to\DSTDST$ events, and $\MMSq$ distinguishes between $\DSTD\to\gamma \DD$ and $\DSTD\to\pi^0 \DD$ hypotheses. For events that are not categorized as $\ee\to \DD$ or $\ee\to \DSTD$, a $D^{*}\to \gamma D$ candidate is searched for from the combination of each $D$ candidate and additional photon candidates. Selections are applied on the reconstructed invariant mass and the recoil mass of the $\Dst$ candidate, defined similarly to Eq.~\eqref{eq:mRec}, to suppress backgrounds. $\DSTDSTEven$ events are isolated from the other $\DSTDST$ decay hypotheses with the variable $\MMSqGamma$, defined similarly to the variable in Eq.~\eqref{eq:kinvars}. Flavor-definite \DODO pairs produced from $\DSTpDSTm\to\pi^+\pi^-\DODO$ are suppressed with vetoes on low-momentum charged pions. It should be noted that the isolation selection requirements rely only on partial reconstruction of the final state to maximize selection efficiencies: $\DSTD$ events are identified by only reconstructing the \DD pair, and only a single $\Dst\to\gamma D$ candidate is reconstructed to identify $\DSTDST$ events. Due to the low efficiency and resolution with which the slow neutral pions from $\Dst\to\pi^0 D$ decays are reconstructed, this decay mode is not pursued and $\DSTDST\to\pi^0\pi^0\DD$ events are incorporated with low efficiency. Further details of the selection criteria are presented in Ref.~\cite{CompanionPaper}.

\begin{figure}[b!]
    \centering
    \includegraphics[width=\linewidth]{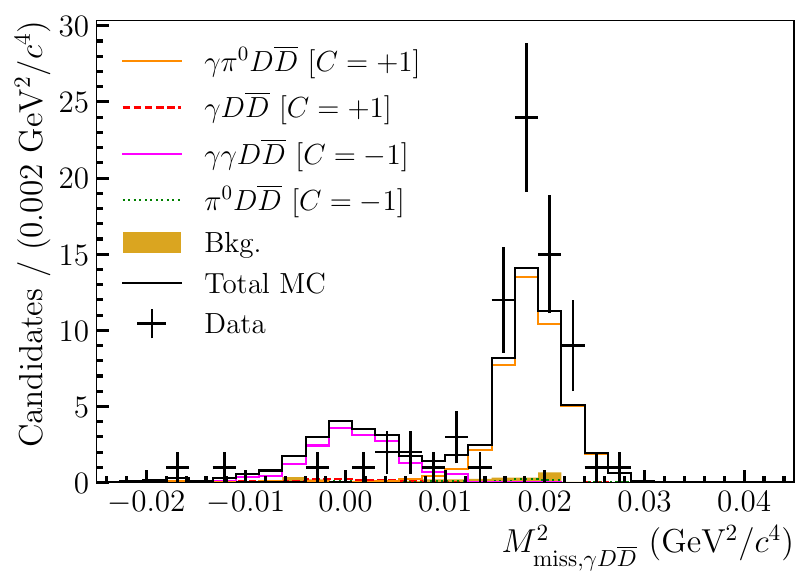}
    \caption{$\MMSqGamma$ distribution of the \mbox{$\DD\to K_S^0\pi^0\text{ vs. }K_S^0\pi^0$} candidates identified with an $\ee\to\DSTDST$ hypothesis in data, compared to a simulation without quantum correlations.}
    \label{fig:SelDist}
\end{figure}

Evidence of quantum correlations can be seen in the distributions of the selection variables, exemplified  in Fig.~\ref{fig:SelDist} for \mbox{$\DD\to K_S^0\pi^0\text{ vs. }K_S^0\pi^0$} decays from $\ee\to \DSTDST$ production, where an enhancement and suppression in the data can be seen in the $\C$-even and \C-odd regions, respectively, relative to the correlation-free simulation.

	




 \begin{figure*}
\includegraphics[width=0.8\linewidth]{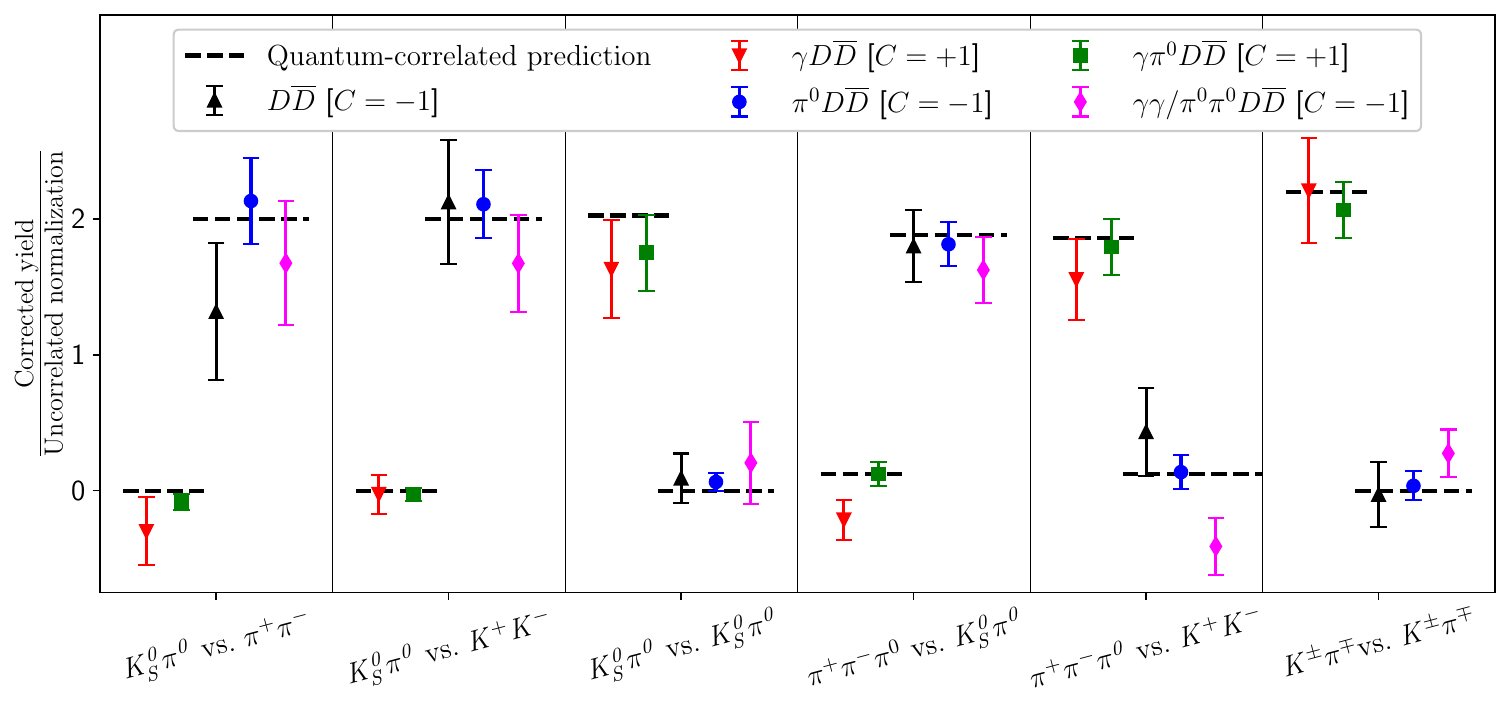}
\caption{The ratios of efficiency-corrected yields observed in data to those expected in the absence of correlations and mixing for each $\DD$ final state originating from each production mechanism. The displayed errors include systematic uncertainties.}
\label{fig:QCDemo}
\end{figure*}

The isolation selection requirements sort correctly reconstructed $\DD$ pairs into production-mechanism hypotheses with high efficiency and purity, but still introduce non-negligible rates of production-mechanism misidentification. This means that the observed yield from a given production-mechanism hypothesis may contain contributions from \DD pairs constrained to a  \C eigenvalue opposite to what is assumed. Production mechanism cross-feed of this sort and selection and reconstruction efficiencies are simultaneously accounted for through an unfolding equation:
 \begin{equation}\label{eq:unfolding}
    \vec{n}= A^{-1} \times \vec{N},
 \end{equation}
 where, for a given $\DD$ final state, $\vec{n}$ gives the efficiency-corrected number of events from each production mechanism, $\vec{N}$ contains the observed yields from each production-mechanism hypothesis, and the entries of the matrix $A_{ij}$ give the probabilities of an event produced from production mechanism $i$ being identified with production-mechanism hypothesis $j$.

 Table~\ref{table:CPDTags} lists the six examined \DD final states whose rates are expected to be maximally affected by the \C constraints. Additional combinations of the examined $\CP$-eigenstate decays are not considered due to limited sample sizes or large backgrounds. The six final states, the \mbox{$\DD\to K^-\pi^+\text{ vs. } K^+\pi^-$} final state considered for normalization, and the five production-mechanism hypotheses for each final state leave thirty-five observed yields which must be determined. This is achieved with fits to the two-dimensional distribution of the reconstructed invariant mass of the $D$ candidates, which determines the observed yields of correctly reconstructed \DD pairs and subtracts the small contributions from backgrounds that mimic the $\DD$ final state. 

\renewcommand{\arraystretch}{1.2}
 \begin{table}[h!]
    \centering
    
    \caption{List of examined \DD final states and their expected $\CP$ eigenvalues. The rates of the first six listed final states are expected to be significantly affected by quantum correlations.}
    \label{table:CPDTags}
    \begin{tabular}{ccc|c}
    \multicolumn{3}{c}{$\DD\to$}  & $\CP$ eigenvalue\footnote{The decay $D\to\pi^+\pi^-\pi^0$ is not exactly $\CP$-definite, but is treated as such with a correction factor based on the measured \CP-even fraction~\cite{pipipiFPlus}.}\\
     \hline
        $\pi^+\pi^-\pi^0$& vs. &$K_S^0\pi^0$ &  $-1$ \\
        $\pi^+\pi^-\pi^0$& vs. &$K^+K^-$&  $+1$\\
        $K_S^0\pi^0$& vs. &$K_S^0\pi^0$ &  $+1$\\
        $K_S^0\pi^0$& vs. &$K^+K^-$ &  $-1$\\
        $K_S^0\pi^0$& vs. &$\pi^+\pi^-$ &  $-1$\\
        $K^\pm\pi^\mp$& vs. &$K^\pm\pi^\mp$  & --\\
        \hline
        $K^-\pi^+$& vs. &$K^+\pi^-$ &  --\\
    \end{tabular}

\end{table}


 The efficiency and cross-feed corrected yields for each of the six final states in Table~\ref{table:CPDTags} are determined based on Eq.~\eqref{eq:unfolding} and expected to be significantly affected by the \C-constraints. The impact of the quantum correlations in each of these final states is studied as a ratio to the number of observed \mbox{$\DD\to  K^-\pi^+\text{ vs. }K^+\pi^-$} decays, which has a  rate expected to be modified by less than 1\% from the \C constraints. For simplicity of comparison, the ratios are additionally scaled by measured branching fractions and efficiencies such that the expected ratios in the absence of \C-correlations and mixing effects are exactly unity. For a production mechanism constrained to an eigenvalue \C, the ratios for \CP-eigenstate $\DD$ pairs are expected to be $0$ if $\C\neq\CP$ and $2$ otherwise~\footnote{These predictions are again modified by a correction factor for final states with a $\pi^+\pi^-\pi^0$ decay~\cite{pipipiFPlus}.}, to linear order in mixing terms. While the $\D\to K^\pm\pi^\mp$ decay is not \CP-definite, the decay $\DD\to K^\pm\pi^\mp\vs K^\pm\pi^\mp$ is forbidden due to the requirement of antisymmetric exchange of the \D and \Db\ mesons in a \C-odd constrained system, while the \C-even ratio is predicted to be $2.2$, due to additional enhancement from mixing effects. The measured ratios are compared to the quantum-correlated predictions in Fig.~\ref{fig:QCDemo}, which clearly demonstrates the presence of quantum correlations in all examined production mechanisms.

 The information on the correlations from both enhanced and forbidden final states is combined by introducing independent coherence factors for each production mechanism, as defined in Eq.~\eqref{eq:kappadef}. The quantum-correlated predictions are reformulated to incorporate the coherence factors, and the coherence factors are then determined by a simultaneous $\chi^2$ fit to the thirty observed yields expected to be forbidden or enhanced. Systematic uncertainties due to imperfect knowledge of reconstruction efficiencies and input branching fractions are accounted for through Gaussian constraints in the fit, but contribute very little to the determined uncertainties. The fit is of good quality, with a $\chi^2$ per degree of freedom of $19.9/25$.  The fitted coherence factors and their uncertainties are shown in Table~\ref{tab:kappas}. The results all show good agreement with the assumption of exact coherence in each of the studied production mechanisms and are all incompatible with the absence of quantum correlations, which validates the procedure for analyzing the interference effects in $\ee\to X\DD$ processes.

\renewcommand{\arraystretch}{1.2}
\begin{table}[ht!]
    \centering
        \caption{    \label{tab:kappas} The coherence factors determined by fit to the observed yields of the processes in Table~\ref{table:CPDTags}. }
    \begin{tabular}{c|c}
        Production mechanism& $\kappa$ \\ \hline
        \DD & $1.015\pm0.066$  \\
        \DSTDG & $1.044\pm0.044$\\
        \DSTDP&$1.028 \pm 0.024$ \\
        \DSTDSTEven & $1.027\pm0.017$ \\
        \DSTDSTOdd & $0.963\pm0.060$
    \end{tabular}

\end{table}


 Given the demonstrated coherence, these samples can be leveraged to measure hadronic parameters. The procedure and results of Ref.~\cite{CompanionPaper} are summarized here, which employs the mixture of $\C$-even and $\C$-odd correlations in a novel measurement technique of \deltaKpi, the strong-phase difference between $\Dz\to K^-\pi^+$ and \mbox{$\Dzb\to K^-\pi^+$} decay amplitudes.  

 The measurement of \deltaKpi with these data proceeds similarly to what is presented in the demonstration of the correlations. Decays of \mbox{$\DD\to K^\mp\pi^\pm\text{ vs. }Y$} are reconstructed, where $Y$ is one of the following recoil $D$ decay modes: $\left\{\pi^+\pi^-\pi^0, K^+K^-, \pi^+\pi^- ,K_S^0\pi^0, K_S^0\pi^+\pi^-\right\}$. The $\DD$ pairs are then sorted into production-mechanism hypotheses using the same selection requirements described in the demonstration of quantum correlations, and observed yields are determined through fits to the reconstructed invariant masses of the two $D$ candidates. 
  
 The examined $\CP$-eigenstate recoil $D$ decays are analyzed to measure $\rKpi\cos\deltaKpi$, where $\rKpi$ is the magnitude of the ratio of the  $\Dz\to K^-\pi^+$ and $\Dzb\to K^-\pi^+$ decay amplitudes. For each of the $\DD\to K^\mp\pi^\pm$ vs. \CP-eigenstate samples, a $\chi^2$ fit is performed to the ratio of observed yields from the two \C-even production mechanisms to the observed yields in the three \C-odd production mechanisms. The production mechanism cross-feed corrections are parameterized in the fit, and the corrected yields relate to $\rKpi\cos\deltaKpi$ through the expected ratio:

 \begin{equation}
R_{CP}=\frac{\left(1-2\lambda y\right)\left(1+(\rKpi)^2+2\lambda\rKpi\cos\deltaKpi\right)}{\left(1+(\rKpi)^2-2\lambda\rKpi\cos\deltaKpi\right)}~,
 \end{equation}

\noindent where $\lambda=\pm 1$ corresponds to the $\CP$ eigenvalue~\footnote{For the $D\to\pi^+\pi^-\pi^0$ decay, $\lambda$ takes the value $2F_+^{\pi\pi\pi^0}-1$, with $F_+^{\pi\pi\pi^0}$ from Ref.~\cite{pipipiFPlus}.} of the recoil $D$ decay. The parameters $y$ and $\left(\rKpi\right)^2$ are fixed to the values from Ref.~\cite{GammaCombo} in the fit, and contribute negligible uncertainty. Averaging the results of the four \CP-eigenstate decays determines \mbox{$\rKpi\cos\deltaKpi=-0.070\pm0.008$}, where the reported uncertainty is statistical.

    



Samples of $\DD\to K^\mp\pi^\pm\text{ vs. }K_S^0\pi^+\pi^-$ decays allow an additional determination of $\rKpi\cos\deltaKpi$, and are also sensitive to $\rKpi\sin\deltaKpi$ due to the nontrivial strong phases of the $D\to K_S^0\pi^+\pi^-$ decay amplitude. This sample is analyzed in bins of the $D\to K_S^0\pi^+\pi^-$ phase space defined by the ``equal-$\Delta\delta_D$" binning scheme from Ref.~\cite{psippKspipiPRD}.  The parameters  $\rKpi\cos\deltaKpi$ and $\rKpi\sin\deltaKpi$ are determined by a simultaneous fit to the reconstructed invariant-mass distributions in each phase-space bin from each production-mechanism hypothesis, which partitions the total sample into 80 subsamples. Cross-feed between phase-space bins and production mechanisms are accounted for in the fit. The $D\to K_S^0\pi^+\pi^-$ strong-phase parameters are fixed to the values determined in Refs.~\cite{psippKspipiPRL,psippKspipiPRD}, and the charm-mixing parameters $x$ and $y$ to the values determined in Ref.~\cite{GammaCombo}.  The fit determines $\rKpi\cos\deltaKpi=-0.044 \pm 0.014$ and $\rKpi\sin\deltaKpi=-0.022 \pm 0.017$, where both uncertainties are only statistical, with a $\chi^2$ per degree of freedom of $85.1/73$.

The results of the \CP-eigenstate and $K_S^0 \pi^+\pi^-$ recoil decays in this work are combined accounting for correlated systematic uncertainties~\footnote{Results as a function of the input parameters $x$, $y$, and $\rKpi$ are presented in Ref.~\cite{CompanionPaper} to allow reinterpretation as their values become better known. } to determine \mbox{$\delta_{K\pi}=\left(192.8^{+11.0 + 1.9}_{-12.4 -2.4}\right)^\circ$},
with the value of \rKpi taken from Ref.~\cite{GammaCombo}. The result agrees well with the most precise measurement from BESIII, performed at the $\psi(3770)$ resonance, $\deltaKpi=\left(187.6^{+8.9 + 5.4}_{-9.7 -6.4}\right)^\circ$~\cite{deltaKpi}, and a recent indirect determination from the LHCb experiment \mbox{$\deltaKpi=(190.2\pm2.8)^\circ$} from Ref.~\cite{GammaCombo}. 

 The presence of both $\C$-even and $\C$-odd correlated samples allows for robust control of systematic uncertainties. This is highlighted in comparison to Refs.~\cite{deltaKpi,BESIII:2014rtm}, which determined $\deltaKpi$ from $\ee\to\DD$ events collected near the production threshold. The examination of the ratios of \CP-eigenstate yields between \C-even and \C-odd correlated samples provides cancellation in all systematic uncertainties associated with reconstructing the $\DD$ final state. For reference, such systematics contributed an uncertainty roughly two-thirds of the statistical component on the determination of $\rKpi\cos\deltaKpi$ from $\CP$-eigenstate decays in Ref.~\cite{deltaKpi}. The mixed \C-constraints also benefit the analysis of $\DD\to K^\mp\pi^\pm \text{ vs. } K_S^0\pi^+\pi^-$ decays, where the leading uncertainties arise from the measurements of hadronic parameters of the $D\to K_S^0\pi^+\pi^-$ decay determined in Ref.~\cite{psippKspipiPRD}. The total systematic uncertainty on the determination of $\deltaKpi$  in this work is reduced by roughly a factor of three compared to Ref.~\cite{deltaKpi}, and arises from production mechanism cross-feed corrections and modeling small background contributions. 

The total precision of this measurement with $7.13\invfb$ of data collected between $\Ecm=4.13-4.23\GeV$ is roughly equivalent to the previous measurement with $2.93\invfb$ of BESIII data collected at $\Ecm=3.77\GeV$ presented in Ref.~\cite{deltaKpi}. The analysis presented in this Letter determines $\deltaKpi$ with a larger statistical uncertainty despite a larger integrated luminosity and comparable inclusive cross-section of $\ee\to \DD$. The larger statistical uncertainty arises from the challenge of $D$ decays with missing particles in the final state, such as $D^0\to K_L^0\pi^+\pi^-$, as the production mechanisms are partially reconstructed to optimize the yields of fully-reconstructed \DD final states, and from the low efficiency with which $\DSTDST\to \pi^0\pi^0 \DD$ decays are reconstructed. Despite this, the analysis presented in this Letter demonstrates that \D meson strong-phase measurements with $\ee$ data collected at higher energy ranges would complement the established approach of studying $\D$-strong phases with data collected near the $\psi(3770)$ energy threshold.

Given the similar sensitivities, the result from this measurement is combined with the result from Ref.~\cite{deltaKpi} accounting for correlated uncertainties due to the assumed hadronic parameters of the $D\to K_S^0\pi^+\pi^-$ decay. The combined analysis gives $\deltaKpi=(189.2^{+6.9+3.4}_{-7.4-3.8})^\circ$.

 
 Measurements of $\C$-even constrained \DD pairs similar to those presented in this Letter with future datasets of $\ee$ collisions collected by BESIII and the proposed super $\tau$-charm facility (STCF)~\cite{STCF} could also constrain the values of charm-mixing parameters, as discussed in Refs.~\cite{Goldhaber:1976fp,Bondar:2010qs,Rama:2015pmr,Cheng:2025kpp}. Based on the procedure presented in this Letter, the sensitivity of STCF to the charm-mixing parameters $x$ and $y$ and the parameters corresponding to \CP-violation in mixing~\cite{LHCbBinFlip}, $\left|\frac{q}{p}\right|$ and  $\phi$, is examined. The sensitivities at several center-of-mass energies primarily proposed for other physics goals are analyzed: $\Ecm=4.01\GeV$, $\Ecm=4.18\GeV$, and $\Ecm=4.26\GeV$. However, center-of-mass energies near the $\psi(4040)$ resonance are expected to be optimal for collection of quantum-correlated $\DD$ pairs due to  large cross-sections of both the $\ee\to \DSTD$ and $\ee\to \DSTDST$ processes, so  the sensitivity from data collected at $\Ecm=4.03\GeV$ is also considered. Assumed cross-sections are taken from the measurements in Ref.~\cite{CLEOXSecs}, and reconstruction efficiencies are estimated based on current BESIII results. All combinations of the decay modes considered in this Letter, the  $D\to K^\mp\pi^\pm\pi^0$ decay, and the $D\to K^\mp\pi^\pm\pi^+\pi^-$ decay using the binned results from Ref.~\cite{BESIIIK3Pi} are included in the sensitivity studies. The best known values for branching fractions~\cite{ParticleDataGroup:2024cfk} and hadronic parameters~\cite{GammaCombo,psippKspipiPRD,BESIIIK3Pi} are assumed, and the uncertainty on these parameters is neglected as sufficient future precision can be achieved with the $\C$-odd samples produced from the same data. The possibility of direct $\CP$-violation is neglected, although it should be noted that the correlated $\C$-odd production mechanisms could be used to probe such effects due to their quadratic dependence on mixing parameters. The precision with the considered decays modes assuming $1\text{ ab}^{-1}$  at each energy point is shown in Table~\ref{table:sensitivity}, which allows for comparison in sensitivity at these energy points. The final distribution of luminosity at STCF should be informed by more detailed studies of the $\ee\to \DSTD$ and $\DSTDST$ production cross-sections near the $\psi(4040)$ resonance and will need to be optimized against competing physics goals. It should be noted that the stated sensitivities depend strongly on the assumed hadronic parameters of the included modes.

 \begin{table}[hbtp]
     \centering
     \caption{Projected statistical sensitivity to charm-mixing parameters with $1\text{ ab}^{-1}$ at various $\ee$ collision energies using the $\D$ decay modes considered in this Letter and  the $D\to K^\mp\pi^{\pm}\pi^0$ and $D\to K^\mp\pi^{\pm}\pi^+\pi^-$  decays. The combined sensitivity with $1\text{ ab}^{-1}$ samples at each energy is also shown.  }
     \label{table:sensitivity}
     \begin{tabular}{c|c c c c}

    $\Ecm\; (\text{GeV})$ & $x\; (10^{-4})$     &  $y\; (10^{-4})$     & $\left|\frac{q}{p}\right| \;(\%)$  &
     $\phi\; ({}^\circ)$ \\
     \hline
$4.01$ & 8.3 & 2.4 & 4.4 & 3.2 \\
$4.03$ &6.9 & 2.0 &  3.4 & 2.5\\ 
$4.18$ & 7.9 & 2.3 & 4.1 & 3.0 \\
$4.26$  & 11.4 & 3.2 & 6.7 & 4.8 \\ \hline
Combined ($4\text{ ab}^{-1}$)  & 5.6 & 1.5 & 2.1 & 1.6
     
     \end{tabular}

 \end{table}

 The results of the sensitivity studies suggest that measurement techniques similar to those presented in this Letter at a high-intensity charm factory such as STCF would provide roughly equivalent sensitivity to the current global averages~\cite{GammaCombo,UTFit} dominated by results from the LHCb collaboration: $4.4\times 10^{-4}$ on $x$, $2.1\times 10^{-4}$ on $y$, $1.8\%$ on $\left|\frac{q}{p}\right|$, and $1.2^\circ$ on $\phi$. This would provide the first confirmation of a non-zero difference of mass of $\Dz$ and $\Dzb$ mesons outside of the LHCb collaboration and independent confirmation of other parameters to current precision from a vastly different measurement technique.  The inclusion of additional decay modes such as  $D\to K^\mp e^{\pm}\nu_e$, considered in Ref.~\cite{Bondar:2010qs}, and $D\to K_S^0  \pi^+ \pi^-\pi^0$, whose strong-phases are not currently well-known, would additionally improve the expected precision on charm-mixing parameters.

 In this Letter, the first demonstration of quantum correlations of $\DD$ pairs produced in \mbox{$\ee\to X \DD $} decays has been presented, including the first observation of $\C$-even correlated $\DD$ pairs. These processes have been employed to measure the strong phase between $D^0\to K^-\pi^+$ and $\Dzb\to K^-\pi^+$ decays, finding \mbox{$\delta_{K\pi}=\left(192.8^{+11.0 + 1.9}_{-12.4 -2.4}\right)^\circ$}. This opens up new prospects for the measurement of $D$-meson hadronic parameters and paves the way towards time-integrated measurements of $\Dz$--\Dzb mixing from $\ee$ colliders.

\input{acknowledgments}

\bibliography{apssamp}



\end{document}

%% file: definitions.tex
\usepackage{xspace} 

\def\beq{\begin{equation}}
\def\eeq{\end{equation}}

\def\CP{$CP$ }
\def\C      {\ensuremath{C}\xspace}

\def\vs{\text{ vs. }}

\def\psipp{\psi(3770)}

\def\DODO{\ensuremath{D^{0}}\ensuremath{\bar{D}^{0}}\xspace}
\def\DSTO{\ensuremath{D^{*0}}\xspace}

\def\aDSTO{\bar{D}^{*0}}

\def\DSTpDSTm{\ensuremath{D^{*+}}\ensuremath{D^{*-}}\xspace}
\def\Dst{D^*}

\def\DODO{\ensuremath{\Dz\Dzb}\xspace}
\def\DD{\ensuremath{\D\Db}\xspace}
\def\DSTDG{\ensuremath{\Dstar\Db\to {\gamma} \D\Db}\xspace}
\def\DSTDP{\ensuremath{\Dstar\Db\to{\pi^0} \D \Db}\xspace}
\def\DSTDSTEven{\ensuremath{\Dstar\Dstarb\to{\gamma\pi^0}\D \Db}\xspace}
\def\DSTDSTOdd{\ensuremath{\Dstar\Dstarb\to{\gamma\gamma/\pi^0\pi^0}\D \Db}\xspace}

\def\DSTD{\ensuremath{\Dstar\Db\xspace}}

\def\DSTDST{\ensuremath{\Dstar\Dstarb}\xspace}

\def\ee{e^+e^-}

\def\Ecm{E_{\text{cm}}}

\def\Emiss{E_{\text{miss}}}
\def\MMSq{M_{\text{miss,$\D\Db$}}^2}
\def\MMSqGamma{M_{\text{miss,$\gamma D\Db$}}^2}
\def\DStarMRec{\Delta M_{\text{rec,}D}}
\newcommand{\MRec}[1]{M_{\text{rec,}#1}}

\def\invfb{\text{ fb}^{-1}}

\def\GeV{\text{ GeV}}

\def\kaon    {{\ensuremath{\PK}}\xspace}

\def\thebaroffset{0.1em}
\newcommand{\offsetoverline}[2][\thebaroffset]{\kern #1\overline{\kern -#1 #2}}%

\def\Ppi         {\ensuremath{\pi}\xspace} 
\def\pion   {{\ensuremath{\Ppi}}\xspace}

\def\pip    {{\ensuremath{\pion^+}}\xspace}

\def\kaon    {{\ensuremath{\textit{K}}}\xspace}

\def\Km      {{\ensuremath{\kaon^-}}\xspace}

\def\PD      {\ensuremath{D}\xspace}  
\def\Dbar    {{\ensuremath{\offsetoverline{\PD}}}\xspace}
\def\D       {{\ensuremath{D}}\xspace}
\def\Db      {{\ensuremath{\offsetoverline{{D}}}\xspace}}
\def\Dz      {{\ensuremath{\D^0}}\xspace}
\def\Dzb     {{\ensuremath{\Dbar{}^0}}\xspace}
\def\Dp      {{\ensuremath{\D^+}}\xspace}
\def\Dm      {{\ensuremath{\D^-}}\xspace}

\def\DpDm    {\ensuremath{\Dp {\kern -0.16em \Dm}}\xspace}
\def\Dstar   {{\ensuremath{\D^*}}\xspace}
\def\Dstarb  {{\ensuremath{\Dbar{}^*}}\xspace}
\def\Dstarz  {\DSTO\xspace}
\def\Dstarzb {\aDSTO\xspace}

\def\DDb     {{\ensuremath{\D \Db}}\xspace}

\def\Ppsi        {\ensuremath{\psi}\xspace}  
\def\psipp  {{\ensuremath{\Ppsi(3770)}}\xspace}

\def\DstarzDstarzb    {\ensuremath{\Dstarz {\kern -0.07em \Dstarzb}}\xspace}

\def\deltaKpi      {\ensuremath{\delta_{K\pi}^\D}\xspace}

\def\C      {\ensuremath{C}\xspace}
\def\CP                {{\ensuremath{C\!P}}\xspace}
\def\rKpi                {{\ensuremath{r_{K\pi}^\D}}\xspace}

\newcommand{\aunit}[1]{\ensuremath{\text{\,#1}}}       
    
\def\fb   {\ensuremath{\aunit{fb}}\xspace}
\def\invfb   {\ensuremath{\fb^{-1}}\xspace}
 


%% file: authorlist_2025-04-06.tex
\author{
\begin{small}
\begin{center}
M.~Ablikim$^{1}$, M.~N.~Achasov$^{4,c}$, P.~Adlarson$^{77}$, X.~C.~Ai$^{82}$, R.~Aliberti$^{36}$, A.~Amoroso$^{76A,76C}$, Q.~An$^{73,59,a}$, Y.~Bai$^{58}$, O.~Bakina$^{37}$, Y.~Ban$^{47,h}$
% ...
\\
\vspace{0.2cm}
(BESIII Collaboration)\\
\vspace{0.2cm} {\it
$^{1}$ Institute of High Energy Physics, Beijing 100049, People's Republic of China\\
% ...
$^{4}$ Budker Institute of Nuclear Physics SB RAS (BINP), Novosibirsk 630090, Russia\\
% ...
$^{36}$ Johannes Gutenberg University of Mainz, Johann-Joachim-Becher-Weg 45, D-55099 Mainz, Germany\\
$^{37}$ Joint Institute for Nuclear Research, 141980 Dubna, Moscow region, Russia\\
% ...
$^{47}$ Peking University, Beijing 100871, People's Republic of China\\
% ...
$^{58}$ Southeast University, Nanjing 211100, People's Republic of China\\
$^{59}$ State Key Laboratory of Particle Detection and Electronics, Beijing 100049, Hefei 230026, People's Republic of China\\
% ...
$^{73}$ University of Science and Technology of China, Hefei 230026, People's Republic of China\\
% ...
$^{76}$ University of Turin and INFN, (A)University of Turin, I-10125, Turin, Italy; (B)University of Eastern Piedmont, I-15121, Alessandria, Italy; (C)INFN, I-10125, Turin, Italy\\
$^{77}$ Uppsala University, Box 516, SE-75120 Uppsala, Sweden\\
% ...
$^{82}$ Zhengzhou University, Zhengzhou 450001, People's Republic of China\\
\vspace{0.2cm}
$^{a}$ Deceased\\
% ...
$^{c}$ Also at the Novosibirsk State University, Novosibirsk, 630090, Russia\\
% ...
$^{h}$ Also at State Key Laboratory of Nuclear Physics and Technology, Peking University, Beijing 100871, People's Republic of China\\
% ...
}\end{center}
\vspace{0.4cm}
\end{small}
}

%% file: acknowledgments.tex
The BESIII Collaboration thanks the staff of BEPCII (https://cstr.cn/31109.02.BEPC) and the IHEP computing center for their strong support. This work is supported in part by National Key R\&D Program of China under Contracts Nos. 2023YFA1606000, 2023YFA1606704; National Natural Science Foundation of China (NSFC) under Contracts Nos. 11635010, 11935015, 11935016, 11935018, 12025502, 12035009, 12035013, 12061131003, 12192260, 12192261, 12192262, 12192263, 12192264, 12192265, 12221005, 12225509, 12235017, 12361141819; the Chinese Academy of Sciences (CAS) Large-Scale Scientific Facility Program; CAS under Contract No. YSBR-101; 100 Talents Program of CAS; The Institute of Nuclear and Particle Physics (INPAC) and Shanghai Key Laboratory for Particle Physics and Cosmology; Agencia Nacional de Investigación y Desarrollo de Chile (ANID), Chile under Contract No. ANID PIA/APOYO AFB230003; ERC under Contract No. 758462; German Research Foundation DFG under Contract No. FOR5327; Istituto Nazionale di Fisica Nucleare, Italy; Knut and Alice Wallenberg Foundation under Contracts Nos. 2021.0174, 2021.0299; Ministry of Development of Turkey under Contract No. DPT2006K-120470; National Research Foundation of Korea under Contract No. NRF-2022R1A2C1092335; National Science and Technology fund of Mongolia; Polish National Science Centre under Contract No. 2024/53/B/ST2/00975; STFC (United Kingdom); Swedish Research Council under Contract No. 2019.04595; U. S. Department of Energy under Contract No. DE-FG02-05ER41374.